\documentclass[aps,prd,showpacs,showkeys,amsmath,amssymb,
twocolumn,
floatfix,
superscriptaddress
]{revtex4-1}
\usepackage{graphicx}
\usepackage{times}
\usepackage{verbatim}
\usepackage{multirow}
\usepackage[usenames,dvipsnames]{color}
\usepackage[normalem]{ulem}
\usepackage{accents}

\newcommand{\pbar}[1]{\accentset{(-)}{#1}}

\newcommand{\ECUST}{\affiliation{Institute of Modern Physics, East China University of Science and Technology, Shanghai}}
\newcommand{\IHEP}{\affiliation{Institute~of~High~Energy~Physics, Beijing}}
\newcommand{\Wisconsin}{\affiliation{University~of~Wisconsin, Madison, Wisconsin 53706, USA}}
\newcommand{\Yale}{\affiliation{Department~of~Physics, Yale~University, New~Haven, Connecticut 06520, USA}}
\newcommand{\BNL}{\affiliation{Brookhaven~National~Laboratory, Upton, New York 11973, USA}}
\newcommand{\NTU}{\affiliation{Department of Physics, National~Taiwan~University, Taipei}}
\newcommand{\NUU}{\affiliation{National~United~University, Miao-Li}}
\newcommand{\Dubna}{\affiliation{Joint~Institute~for~Nuclear~Research, Dubna, Moscow~Region}}
\newcommand{\CalTech}{\affiliation{California~Institute~of~Technology, Pasadena, California 91125, USA}}
\newcommand{\CUHK}{\affiliation{Chinese~University~of~Hong~Kong, Hong~Kong}}
\newcommand{\NCTU}{\affiliation{Institute~of~Physics, National~Chiao-Tung~University, Hsinchu}}
\newcommand{\NJU}{\affiliation{Nanjing~University, Nanjing}}
\newcommand{\TsingHua}{\affiliation{Department~of~Engineering~Physics, Tsinghua~University, Beijing}}
\newcommand{\SZU}{\affiliation{Shenzhen~University, Shenzhen}}
\newcommand{\NCEPU}{\affiliation{North~China~Electric~Power~University, Beijing}}
\newcommand{\Siena}{\affiliation{Siena~College, Loudonville, New York  12211, USA}}
\newcommand{\IIT}{\affiliation{Department of Physics, Illinois~Institute~of~Technology, Chicago, Illinois  60616, USA}}
\newcommand{\LBNL}{\affiliation{Lawrence~Berkeley~National~Laboratory, Berkeley, California 94720, USA}}
\newcommand{\UIUC}{\affiliation{Department of Physics, University~of~Illinois~at~Urbana-Champaign, Urbana, Illinois 61801, USA}}
\newcommand{\RPI}{\affiliation{Department~of~Physics, Applied~Physics, and~Astronomy, Rensselaer~Polytechnic~Institute, Troy, New~York  12180, USA}}
\newcommand{\SJTU}{\affiliation{Department of Physics and Astronomy, Shanghai Jiao Tong University, Shanghai Laboratory for Particle Physics and Cosmology, Shanghai}}
\newcommand{\BNU}{\affiliation{Beijing~Normal~University, Beijing}}
\newcommand{\WM}{\affiliation{College~of~William~and~Mary, Williamsburg, Virginia  23187, USA}}
\newcommand{\Princeton}{\affiliation{Joseph Henry Laboratories, Princeton~University, Princeton, New~Jersey 08544, USA}}
\newcommand{\VirginiaTech}{\affiliation{Center for Neutrino Physics, Virginia~Tech, Blacksburg, Virginia  24061, USA}}
\newcommand{\CIAE}{\affiliation{China~Institute~of~Atomic~Energy, Beijing}}
\newcommand{\SDU}{\affiliation{Shandong~University, Jinan}}
\newcommand{\NanKai}{\affiliation{School of Physics, Nankai~University, Tianjin}}
\newcommand{\UC}{\affiliation{Department of Physics, University~of~Cincinnati, Cincinnati, Ohio 45221, USA}}
\newcommand{\DGUT}{\affiliation{Dongguan~University~of~Technology, Dongguan}}
\newcommand{\XJTU}{\affiliation{Xi'an Jiaotong University, Xi'an}}
\newcommand{\UCB}{\affiliation{Department of Physics, University~of~California, Berkeley, California  94720, USA}}
\newcommand{\HKU}{\affiliation{Department of Physics, The~University~of~Hong~Kong, Pokfulam, Hong~Kong}}
\newcommand{\UH}{\affiliation{Department of Physics, University~of~Houston, Houston, Texas  77204, USA}}
\newcommand{\Charles}{\affiliation{Charles~University, Faculty~of~Mathematics~and~Physics, Prague, Czech~Republic}} 
\newcommand{\USTC}{\affiliation{University~of~Science~and~Technology~of~China, Hefei}}
\newcommand{\TempleUniversity}{\affiliation{Department~of~Physics, College~of~Science~and~Technology, Temple~University, Philadelphia, Pennsylvania  19122, USA}}
\newcommand{\CUC}{\affiliation{Instituto de F\'isica, Pontificia Universidad Cat\'olica de Chile, Santiago, Chile}} 
\newcommand{\CGNPG}{\affiliation{China General Nuclear Power Group, Shenzhen}}
\newcommand{\NUDT}{\affiliation{College of Electronic Science and Engineering, National University of Defense Technology, Changsha}} 
\newcommand{\IowaState}{\affiliation{Iowa~State~University, Ames, Iowa  50011, USA}}
\newcommand{\ZSU}{\affiliation{Sun Yat-Sen (Zhongshan) University, Guangzhou}}
\newcommand{\CQU}{\affiliation{Chongqing University, Chongqing}} 
\newcommand{\BCC}{\altaffiliation[Now at ]{Department of Chemistry and Chemical Technology, Bronx Community College, Bronx, New York  10453, USA}} 

\begin{document}

\title{Improved Search for a Light Sterile Neutrino with the Full Configuration of the Daya Bay Experiment}

\author{F.~P.~An}\ECUST
\author{A.~B.~Balantekin}\Wisconsin
\author{H.~R.~Band}\Yale
\author{M.~Bishai}\BNL
\author{S.~Blyth}\NTU\NUU
\author{D.~Cao}\NJU
\author{G.~F.~Cao}\IHEP
\author{J.~Cao}\IHEP
\author{W.~R.~Cen}\IHEP
\author{Y.~L.~Chan}\CUHK
% ...

\collaboration{The Daya Bay Collaboration}\noaffiliation
\date{\today}

\begin{abstract}
This Letter reports an improved search for light sterile neutrino mixing 
in the electron antineutrino disappearance channel 
with the full configuration 
of the Daya Bay Reactor Neutrino Experiment. 
With an additional 404 days of data collected in eight antineutrino detectors, 
this search benefits from 3.6 times the statistics
available to the previous publication, 
as well as from improvements in energy calibration 
and background reduction.
A relative comparison of the rate and energy spectrum of 
reactor antineutrinos in the three experimental halls 
yields no evidence of sterile neutrino mixing in the  
$2\times10^{-4} \lesssim |\Delta m^{2}_{41}| \lesssim 0.3$~eV$^{2}$ mass range. 
The resulting limits on $\sin^{2}2\theta_{14}$ are improved by approximately a factor of two over previous results and constitute the most stringent constraints to date in the $|\Delta m^{2}_{41}| \lesssim 0.2$ eV$^{2}$ region. 
\end{abstract}

\pacs{14.60.Pq, 14.60.St, 28.50.Hw, 29.40.Mc}
\keywords{sterile neutrino, neutrino mixing, reactor neutrino, Daya Bay}
\maketitle

The three-neutrino mixing framework, in which the flavor
  eigenstates ($\nu_e, \nu_\mu, \nu_\tau$) mix with the mass
  eigenstates ($\nu_1, \nu_2, \nu_3$) via the PMNS
  matrix~\cite{Pontecorvo:1957cp, Pontecorvo:1967fh, Maki:1962mu}, has
  been extremely successful in explaining the results observed in most
  solar, atmospheric, reactor and long-baseline accelerator
  neutrino oscillation experiments~\cite{*[] [{, Section 14.}] Agashe:2014kda}. 
Despite this success, the exact mechanism by which neutrinos acquire their mass remains unknown, and the possible existence of additional neutrinos is under active consideration. 
To be consistent with precision electroweak measurements~\cite{ALEPH:2005ab}, these
additional neutrinos are called ``sterile''~\cite{Pontecorvo:1967fh}, that is, noninteracting within the standard model and thus with no known mechanism for direct detection. Nonetheless, an unambiguous signal of their existence can be sought in neutrino oscillation experiments, where they could affect the way in which the three active neutrinos oscillate if they mix with the latter.

In the simplest extension of the standard model, where only one sterile neutrino is considered 
in addition to the three active ones, the mixing can be expressed as
\begin{equation}
\nu_{\alpha} = \sum^{4}_{i=1}U_{\alpha i}\nu_{i},
\end{equation}
where $U$ is a unitary $4 \times 4$ mixing matrix and $U_{\alpha i}$ is the neutrino mixing matrix element for the flavor eigenstate $\nu_\alpha$ and the mass eigenstate $\nu_i$. The survival probability for a relativistic $\nu_\alpha$ with an energy $E$ and a traveling distance $L$ is given by
\begin{equation}
P_{\nu_\alpha \rightarrow \nu_\alpha} = 1 - 4
\sum^{3}_{i=1}\sum^{4}_{j > i}|U_{\alpha i}|^{2}|U_{\alpha j}|^{2}\sin^{2}\Delta_{ji},
\label{eq:psurv-4nu}
\end{equation}
\noindent where $\Delta_{ji}=1.267 {\Delta}m^2_{ji}({\rm eV}^2)\frac{L({\rm m})}{E({\rm MeV})}$ and
${\Delta}m^2_{ji} = m_{j}^2 - m_{i}^{2}$ is the mass-squared difference between the mass eigenstates $\nu_j$ and $\nu_i$.
As indicated in Ref.~\cite{Palazzo:2013bsa}, in the case of electron antineutrino disappearance the neutrino mixing matrix elements $U_{ei}$ can be parametrized in terms of the $\theta_{14}$, $\theta_{13}$ and $\theta_{12}$ mixing angles. Compared with standard three-neutrino mixing, the neutrino oscillation probability includes three additional oscillation frequencies associated with ${\Delta}m^2_{41}, {\Delta}m^2_{42}$ and ${\Delta}m^2_{43}$. 
When $|{\Delta}m^2_{41}| \gg |{\Delta}m^2_{31}|$ these three parameters are virtually indistinguishable, and for the Daya Bay baselines Eq.~(\ref{eq:psurv-4nu}) approximates to
\begin{eqnarray}
  P_{\overline{\nu}_e \rightarrow \overline{\nu}_e}
  \approx 1 &-& 4(1-|U_{e4}|)^2|U_{e4}|^2\sin^{2}\Delta_{41} \nonumber \\
  &-& 4 (1-|U_{e3}|^2-|U_{e4}|^2)|U_{e3}|^2\sin^{2}\Delta_{31} \nonumber \\
  \approx 1 &-& \sin^{2}2\theta_{14}\sin^{2}\Delta_{41}
  - \sin^{2}2\theta_{13}\sin^{2}\Delta_{31}. \quad \quad
\label{eq:psurv-4nu-sim}
\end{eqnarray}
Thus, to first order, 
evidence for light sterile neutrino mixing 
consists of an additional spectral distortion 
with a frequency different from standard three-neutrino oscillations.

\begin{figure}[!htb]
\centering
\includegraphics[width=0.9\columnwidth]{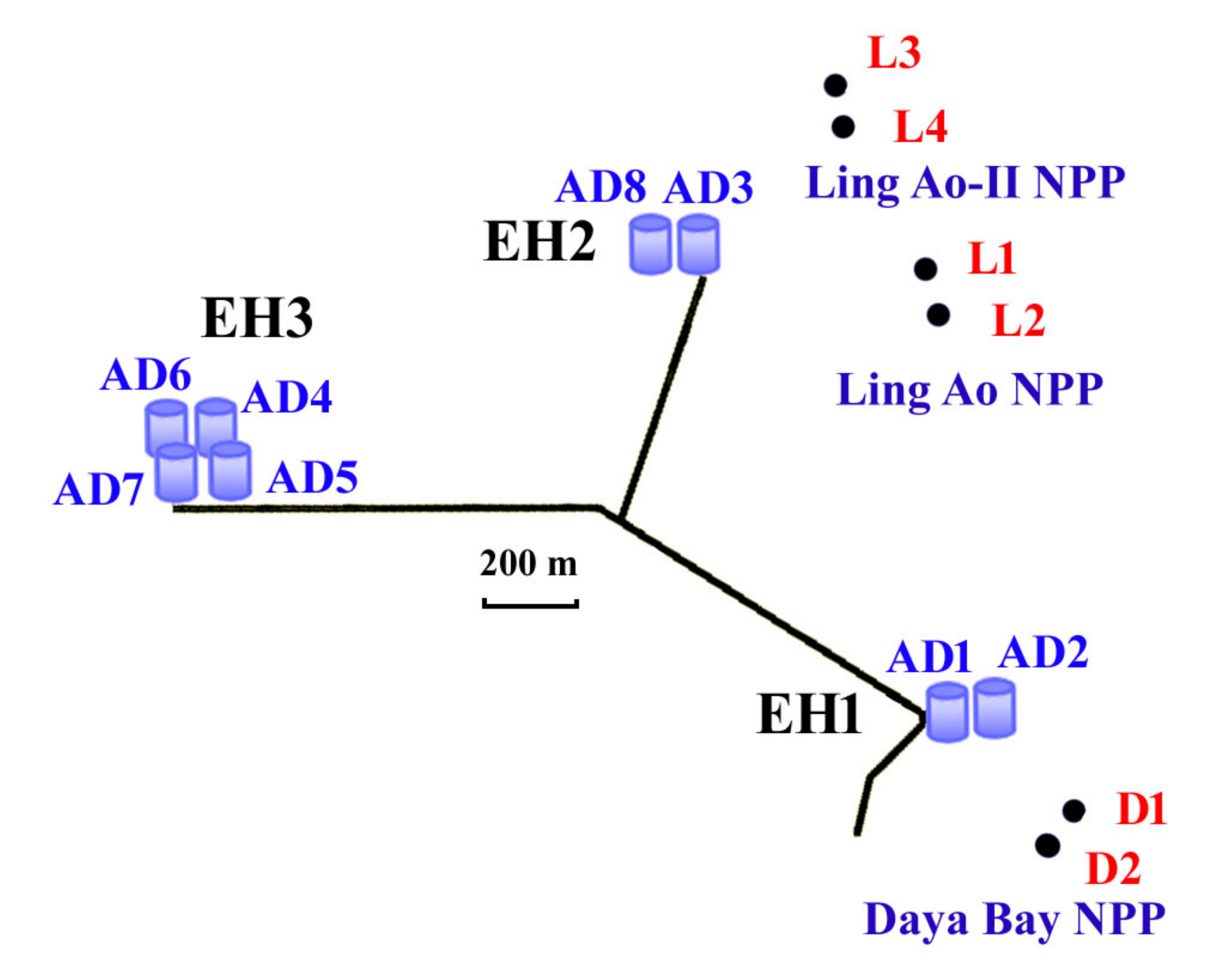}
\caption{
Layout of the Daya Bay Reactor Neutrino Experiment.
The dots represent reactor cores, labeled as D1, D2, L1, L2, L3, and L4. The
Daya Bay experiment started data taking with six antineutrino detectors (AD1-AD6) 
installed in three experimental halls (EH1-EH3). 
From August to October 2012, two additional detectors (AD7 and AD8) 
were installed in EH3 and EH2, respectively.}
\label{fig:layout}
\end{figure}

No conclusive evidence for the existence of sterile neutrinos has been obtained. A few anomalies in short baseline neutrino oscillation experiments \cite{Aguilar:2001ty,Aguilar-Arevalo:2013pmq, AguilarArevalo:2007it,anom,Hampel:1997fc,Abdurashitov:2009tn,Giunti:2010zu} can be explained with additional sterile neutrinos, but these results are in tension with the limits obtained from other experiments~\cite{Karagiorgi:2009nb,Kopp:2011qd, Maltoni:2007zf, Gariazzo:2015rra}. The majority of experimental searches have centered on mass-squared differences around 1 eV$^2$ and higher, whereas the Daya Bay and other medium baseline reactor antineutrino experiments can make unique contributions 
in the sub-eV scale~\cite{Bandyopadhyay:2007rj, deGouvea:2008qk, Bora:2012pi, Kang:2013zma, Bakhti:2013ora, Palazzo:2013bsa, Esmaili:2013yea, Bergevin:2013nea, Bergevin:2013nea, Girardi:2014wea}.
In 2014, the Daya Bay Collaboration reported on a search for light sterile neutrino mixing based on the first 217 days of data acquired with a partial configuration of six functionally identical antineutrino detectors (ADs) deployed at three experimental halls (EHs), the results of which excluded a large, previously unexplored region of parameter space in the $3\times10^{-4} \lesssim |\Delta m^{2}_{41}| \lesssim 0.1$~eV$^{2}$ range ~\cite{An:2014Sterile}. In this partial configuration, three ADs were installed in two near halls (two in EH1 and one in EH2) and another three in a far hall (EH3). This Letter reports on an improved search made with the full eight-detector configuration shown in Figure~\ref{fig:layout} that resulted from the installation of two additional ADs, one in EH2 and another in EH3, in the summer of 2012. The additional 404 days of eight-detector data collected from October 2012 to November 2013 amount to a 3.6 times increase in statistics.

Each AD is a three-zone cylindrical detector composed of
two nested acrylic vessels within a concentric stainless steel
vessel. The central vessel is filled with 20 tons of
  gadolinium-doped liquid scintillator (Gd-LS) that serves as the primary
  target for antineutrino detection. A 22-ton pure
  LS volume encloses the central target and enables detection of $\gamma$ rays that escape from the Gd-LS volume. The outermost cylinder contains 40
  tons of mineral oil that provide shielding against $\gamma$-ray
  radiation from the detector components. 
A total of 192 photomultiplier tubes are installed on
  the vertical surfaces, and the top and bottom surfaces are covered
  with optical reflectors. Three automated calibration units~\cite{Liu:2013ava} that store and deploy calibration sources and LEDs sit on top of the stainless steel vessel.
The ADs are housed inside a muon veto system consisting of two optically separated inner and outer water pools~\cite{Dayabay:2014vka} that provide shielding from ambient radiation 
and serve as active water Cherenkov muon detectors. Four layers of resistive plate chambers are installed on top of each water pool. More information on the Daya Bay detectors and their performance can be found in Refs.~\cite{DayaBay:2012aa,An:2015qga}.

Reactor antineutrinos are detected via the inverse beta decay (IBD) reaction, $\bar\nu_e+p\rightarrow e^+ + n$. 
The positron deposits its energy in the scintillator and 
then annihilates with an electron.
This generates a prompt signal that can be measured with a resolution of $\sigma_{E}/E \sim 8\%$ at 1 MeV and which preserves most of the incident antineutrino's energy.
The neutron is primarily captured by the gadolinium inside the central target, yielding an $\sim$8 MeV delayed signal. 
Requiring coincidence of the prompt and delayed signal pair effectively suppresses backgrounds. 

A summary of the IBD candidates for the six-AD and eight-AD periods, together with the estimated background levels and the baselines of the three experimental halls to each pair of reactor cores, is shown in Table~\ref{tab:ibd}. In the eight-AD period the backgrounds amount to only 2\% of the total candidate samples in the near and far halls~\cite{An:2015rpe}. Two out of three Am-C calibration sources in the automated calibration units on the top of each far AD were removed during the installation of the two additional ADs in the summer of 2012, which reduced the far hall's Am-C background by a factor of 4 compared to that in the previous publication.
This data set also incorporates a reduction in the AD-uncorrelated energy scale uncertainty from $0.35\%$ to $0.2\%$ due to the implementation of better vertex- and time-dependent corrections~\cite{An:2015rpe}. This is one of the dominant systematic uncertainties, and is quantified by studying the differences in detector response using various calibration and natural radioactive sources.

\begin{table*}[!htb]
  \centering
  \caption{Summary of total number of IBD candidates and backgrounds, and baselines of the three experimental halls to the reactor pairs. Statistical and systematic errors are included. 
    \label{tab:ibd}}
  \begin{ruledtabular}
    \begin{tabular}[c]{lccccccc} 
      \multirow{2}*{Site} 
      & \multicolumn{2}{c}{IBD candidates} 
      & \multicolumn{2}{c}{Backgrounds} 
      & \multicolumn{3}{c}{Mean Distance to Reactor Cores (m)} \\
      \cline{2-3}  \cline{4-5} \cline{6-8}  
      & (six-AD) & (eight-AD) 
      & (six-AD) & (eight-AD) 
               &  Daya Bay & Ling Ao &  Ling Ao-II \\\hline
      EH1 & 205135 & 408678 & 4076.6 $\pm$ 462.4 & 7547.9 $\pm$ 908.0 & 365 & 860 & 1310  \\
      EH2 & 93742  & 383402 & 1580.3 $\pm$ 147.8 & 5791.2 $\pm$ 586.5 & 1348 & 481 & 529  \\
      EH3 & 41348  & 108907 & 1878.9 $\pm$ 94.6  & 2105.2 $\pm$ 208.1 & 1909 & 1537 & 1542  \\ 
    \end{tabular}
  \end{ruledtabular}
\end{table*}

\begin{figure}[!htb]
\centering
\includegraphics[width=\columnwidth]{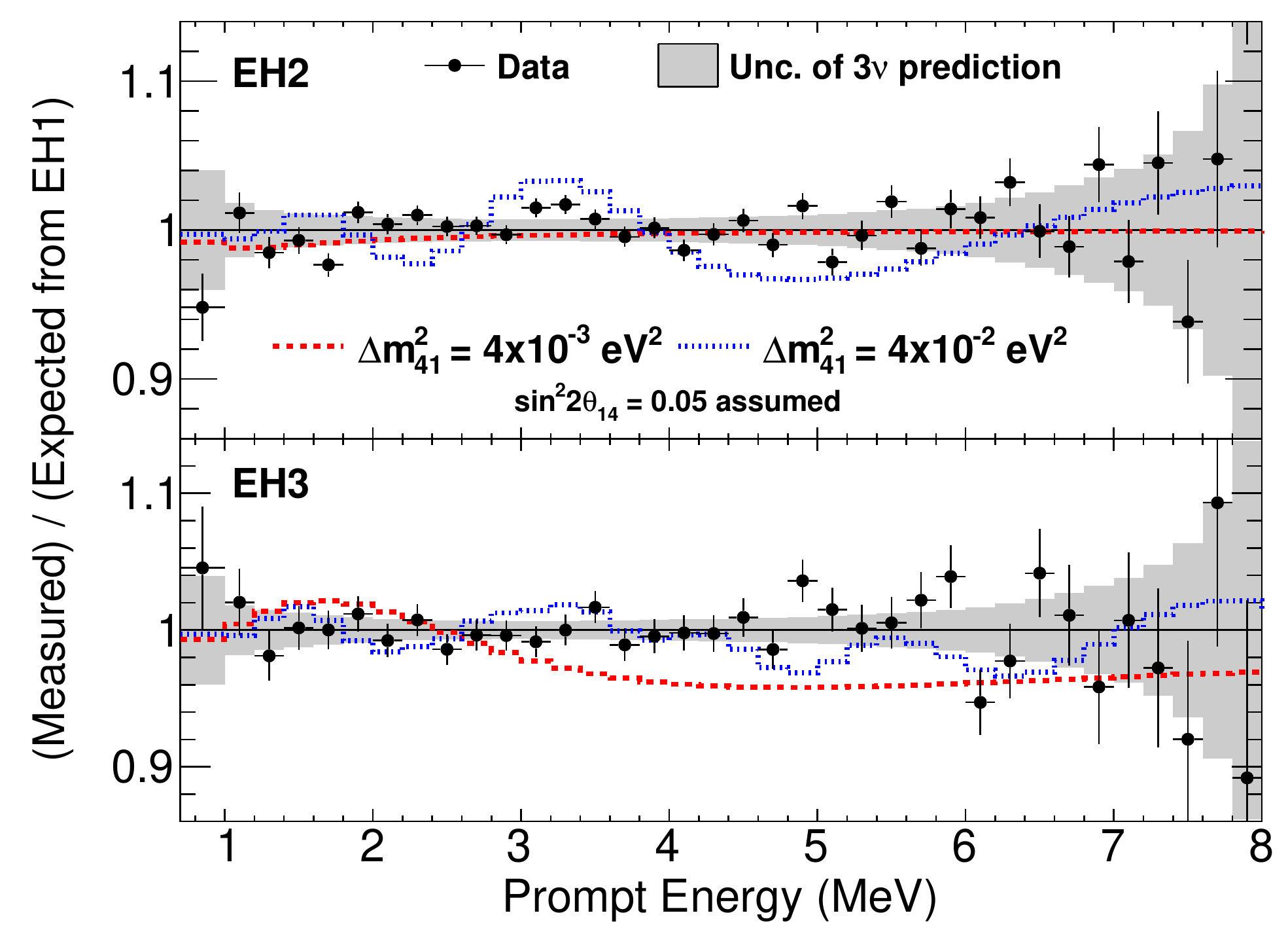}
\caption{Prompt energy spectra observed at EH2 (top)
  and EH3 (bottom), divided by the prediction from EH1
  with the three-neutrino best fit oscillation parameters from the
  most recent Daya Bay analysis~\cite{An:2015rpe}.
The gray band represents the one-standard-deviation uncertainty of the three-neutrino oscillation prediction, which includes the statistical uncertainty of the EH1 data, as well as all the systematic uncertainties. Predictions with $\sin^{2}2\theta_{14} = 0.05$ and two representative $\Delta m^2_{41}$ values are also shown as the dotted and dashed curves.}
\label{fig:SpectralRatio}
\end{figure}

The search for sterile neutrino mixing at the Daya Bay Reactor Neutrino Experiment is carried out through a relative comparison of the antineutrino rates and energy spectra at the three experimental halls. The unique configuration of multiple baselines to three pairs of nuclear reactors allows exploration of $\Delta m^{2}_{41}$ spanning more than 3 orders of magnitude.
Figure~\ref{fig:SpectralRatio} shows the ratios of the observed prompt
energy spectra at EH2 and EH3 to the best fit prediction from EH1 in the three-neutrino case. 
In this figure, the data are compared with the four-neutrino mixing scenario
assuming $\sin^{2}2\theta_{14}=0.05$ for two representative $\Delta m^{2}_{41}$ 
values, illustrating that the sensitivity at $\Delta m^{2}_{41} = 4 \times 
10^{-2} (4 \times 10^{-3})$ eV$^{2}$ originates primarily from the relative spectral 
shape comparison between EH1 and EH2 (EH3). The physical size of the Daya Bay reactor cores and detectors, as well
as the nonuniform distribution of the fission isotopes inside the cores,  have a 
negligible impact on the sensitivity.

The two different analysis methods used in the previous 
search~\cite{An:2014Sterile} were
updated to include the eight-AD data sample. Both methods, referred to as method A and method B, 
use the full expression in Eq.~(\ref{eq:psurv-4nu}) to predict the neutrino oscillation signatures.
The oscillation parameters $\sin^{2}2\theta_{14}$, $\sin^{2}2\theta_{13}$ and $|\Delta m_{41}^{2}|$ are set as free variables,
while the others are constrained through external measurements: $\sin^{2}2\theta_{12} = 0.846 \pm 0.021$, 
$\Delta m_{21}^{2} = (7.53 \pm 0.18)\times 10^{-5}~{\rm eV}^{2}$,
and $|\Delta m_{32}^{2}| = (2.44 \pm 0.06)\times 10^{-3}~{\rm eV^{2}}$~\cite{Agashe:2014kda_all}. 
The normal mass ordering is assumed for both $\Delta m_{31}^{2}$ and $\Delta m_{41}^{2}$, although this choice has only a marginal impact on the results.

Method A explicitly minimizes the dependence on the reactor antineutrino flux modeling~\cite{An:2015rpe} by predicting the prompt energy spectrum at the far hall from the measured spectra at the near halls. This process 
is done independently for each prompt energy bin $i$, 
by applying a weighting factor 
$w_i (\Delta m^2_{41},\sin^22\theta_{14},\sin^22\theta_{13})$
calculated from the known baselines and the reactor power profiles.  
The oscillation hypothesis is tested by evaluating a $\chi^2$ defined as
\begin{equation}
  \label{eq:chi2}
  \chi^{2} =  \sum_{i,j}(N_j^{f} - w_j \cdot
  N_j^{n}) (V^{-1})_{ij} (N_i^{f} - w_i \cdot 
  N_i^{n}), 
\end{equation}
where $N_i^{f(n)}$ is the observed number of events after background
subtraction in the $i$-th bin at a far (near) detector, and
$V$ is a covariance matrix including both systematic and
statistical uncertainties. 
The sensitivity to a spectral distortion between the two near sites is
retained by treating their data separately and by having
indices $i,j$ run over both the EH3-EH1 and
EH3-EH2 combinations. 
A $\chi^2$ constructed with an alternative combination of the near and far detectors, such as EH2-EH1 and EH3-EH1, yields an equivalent sensitivity. All the sources of systematic error included in the most recent
oscillation analysis of Ref.~\cite{An:2015rpe} are considered, in
addition to the uncertainty in the estimation of $\Delta m^2_{32}$.

Method B simultaneously fits the spectra from all ADs
using the predicted reactor antineutrino flux.
A binned log-likelihood function is constructed with nuisance parameters 
for the various systematic terms, 
including the detector response and the backgrounds.  
The reactor antineutrino flux is constrained based on the 
Huber~\cite{Huber:2011wv} and Mueller~\cite{Mueller:2011nm} fissile antineutrino models. 
The spectral uncertainties in the models are enlarged as motivated by the observed discrepancy between the predicted reactor antineutrino spectrum and the data \cite{An:2015nua, An:2016srz, RENO:2015ksa,Abe:2014bwa}, as well as by the recent reexamination of the systematic uncertainties in Ref.~\cite{Hayes:2013wra}. Specifically, the uncorrelated spectral uncertainties for $^{235}$U, $^{239}$Pu, and $^{241}$Pu are conservatively increased to above 4\%, while that of $^{238}$U is kept above 10\%. The uncertainty of the predicted reactor $\overline{\nu}_{e}$ rate is also increased to 5\%. 

The two complementary analysis methods 
produce practically identical sensitivities for
  $|\Delta m^2_{41}| \lesssim 0.3~\rm eV^2$. 
Method A is more
  robust against uncertainties in the predicted reactor antineutrino
  flux, while method B has a slightly higher reach in sensitivity for $|\Delta m^2_{41}| \gtrsim 0.3~\rm eV^2$ as a result of its incorporation of absolute reactor antineutrino flux constraints. 
The different treatments of systematic uncertainties provide a
  thorough cross-check of the results.
 For method A, the minimum $\chi^2$ value obtained
 with a free-floating $\Delta m^2_{41}$,
$\sin^22\theta_{14}$ and $\sin^22\theta_{13}$ is $\chi_{4\nu}^{2} / {\rm NDF} = 129.1 / 145$,
 where NDF stands for the number of degrees of freedom. The corresponding value in the three-neutrino
  scenario, in which $\sin^22\theta_{13}$ is the only free parameter, is $\chi_{3\nu}^{2}
  / {\rm NDF} = 134.7 / 147$.
  The p-value of observing $\Delta \chi^{2} = \chi_{3\nu}^{2} -
  \chi_{4\nu}^{2} = 5.6$ without sterile neutrino mixing is
  determined to be 0.41 using a large sample of Monte Carlo pseudo-experiments.
  Similarly, the minimum $\chi^2$ values for method B are 
  $\chi_{4\nu}^{2} / {\rm NDF} = 179.74 / 205$ and 
  $\chi_{3\nu}^{2}  / {\rm NDF} = 183.87 / 207$, with a corresponding p-value of 
0.42. As indicated by these p-values, 
  no apparent signature for sterile neutrino mixing is observed. 

The limits in the $(|\Delta
  m^{2}_{41}|,\sin^{2}2\theta_{14})$ plane are also set by two
  independent approaches, the first of which follows the
  Feldman-Cousins method~\cite{Feldman:1997qc}. 
For each point $\boldsymbol{\eta} \equiv (|\Delta
m^{2}_{41}|,\sin^{2}2\theta_{14})$, the value of 
$\Delta\chi^2(\boldsymbol{\eta}) = \chi^2(\boldsymbol{\eta})-
\chi^2(\boldsymbol{\eta}_{\rm best})$ is evaluated,
where $\chi^2(\boldsymbol{\eta})$ is
the smallest $\chi^2$ value with a free-floating
$\sin^22\theta_{13}$.
This $\Delta\chi^2(\boldsymbol{\eta})$ is then compared with 
the critical value
$\Delta\chi^2_{c}(\boldsymbol{\eta})$ encompassing a fraction $\alpha$
of the events, estimated by 
fitting a large number of pseudo-experiments that
include statistical and systematic fluctuations. 
The point $\boldsymbol{\eta}$ is then
declared to be inside the $\alpha$ confidence level (C.L.) acceptance region if
$\Delta\chi^2_{\rm data}(\boldsymbol{\eta}) <
\Delta\chi^2_{c}(\boldsymbol{\eta}).$

The second approach to set the limits is the ${\rm CL_{s}}$ statistical method~\cite{Read:2002hq, CLsMethod}.
For each point in the ($\sin^{2}2\theta_{14}$, $|\Delta m^{2}_{41}|$)
parameter space, a two-hypothesis test is performed in which the null
hypothesis $H_{0}$ is the three-neutrino model and the alternative
hypothesis $H_{1}$ is the four-neutrino model with fixed
$\sin^{2}2\theta_{14}$ and $|\Delta m^{2}_{41}|$. 
The ${\rm CL_{s}}$ value is defined as  
\begin{equation}
{\rm CL_{s}} = \frac{1-p_{1}}{1-p_{0}},
\end{equation}
where $p_{0}$ and $p_{1}$ are the p-values for the three-neutrino and 4-neutrino hypotheses respectively. 
These p-values are calculated from the $\chi^2$ difference of those two 
hypotheses. The value of $\sin^{2}{2\theta_{13}}$ is independently set for each hypothesis based on a fit to the data.
The condition of ${\rm CL_{s}} \leq 1-\alpha$ is required to set 
the ${\rm CL_{s}}$ exclusion region at $\alpha$ confidence level.

  When used with the same analysis method (method A or method B), the difference in sensitivity between the Feldman-Cousins and CL$_\mathrm{s}$ approaches is found to be smaller than 10\%.
  The Feldman-Cousins approach provides a unified method to
  define confidence intervals, but has the drawback that it involves fitting a large amount of
  simulated data sets. Hence, it is used only for method A, which
  eliminates all of the nuisance parameters by utilizing a covariance
  matrix. 
  In contrast, the CL$_\mathrm{s}$ implementation is significantly less computationally intensive, and also provides an alternative for combining the results between multiple experiments~\cite{Read:2002hq, CLsMethod}.
  Accordingly, both the Feldman-Cousins limit from method A and 
  the CL$_\mathrm{s}$ limit from method B are presented in this work. 

\begin{figure}[htb]
\includegraphics[width=\columnwidth]{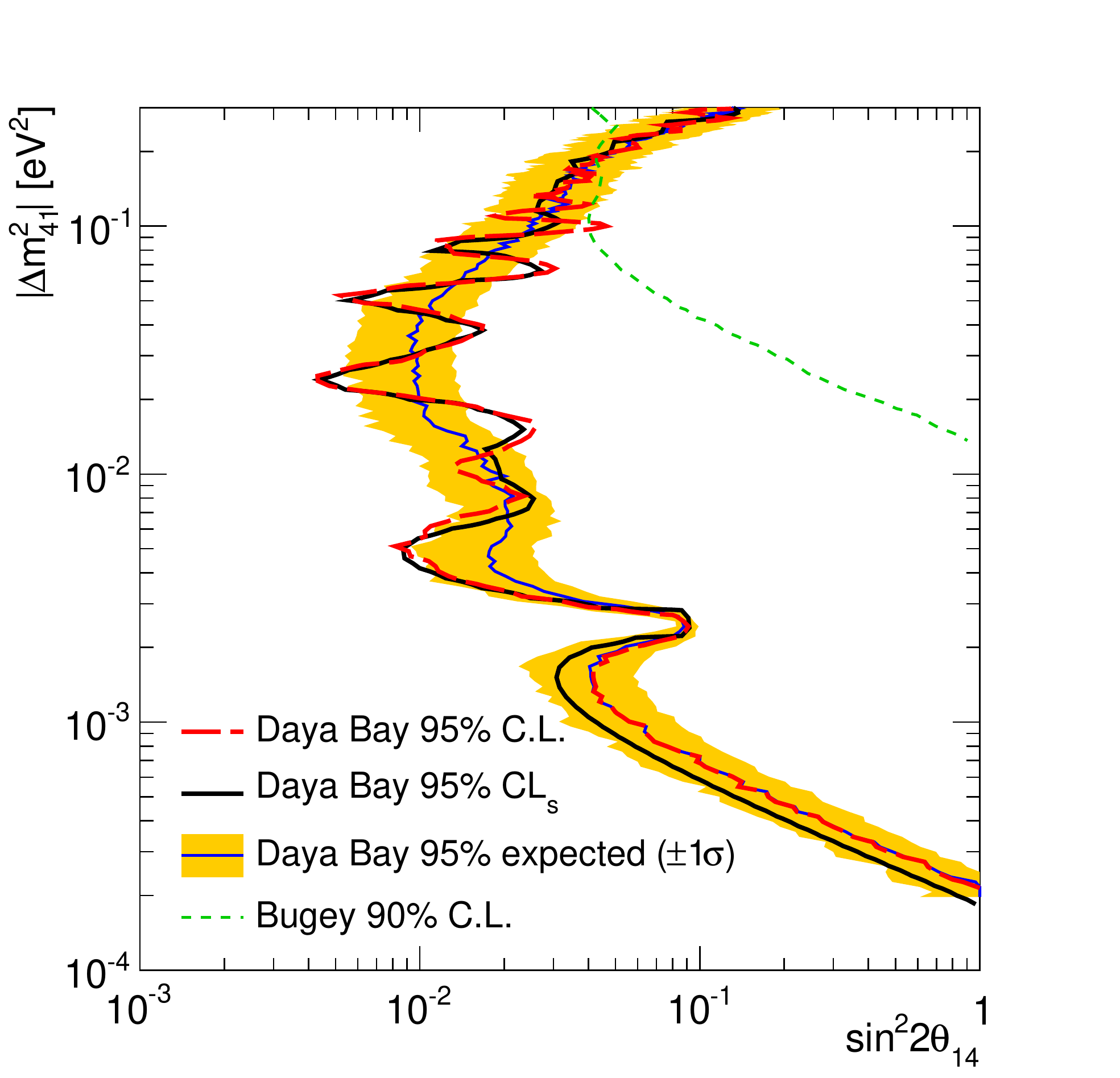}
\caption{
  Exclusion contours in the 
  ($\sin^{2}2\theta_{14}$, $|{\Delta}m^{2}_{41}|$) plane,
 under the assumption of $\Delta m_{32}^{2}>0$ and $\Delta m_{41}^{2}>0$.
  The red long-dashed curve represents the 95\% CL
  exclusion contour with the Feldman-Cousins method~\cite{Feldman:1997qc} from method A.
  The black solid curve represents the 95\% ${\rm CL_{s}}$ exclusion
  contour~\cite{Read:2002hq} from method B.
  The expected 95\% CL 1$\sigma$ band in yellow is centered around the sensitivity curve, shown as a thin blue line. 
  The region of parameter space to
  the right side of the contours is excluded.
  For comparison,
  Bugey's~\cite{Declais:1994su} 90\% CL limit on $\overline{\nu}_e$ disappearance
  is also shown as the green dashed curve. 
  \label{fig:excludeOsc}}
\end{figure}

Figure~\ref{fig:excludeOsc} shows the 95\% confidence level contour
from the Feldman-Cousins approach and the 95\% ${\rm CL_{s}}$ exclusion contour.
Both contours are centered around the 95\% CL expectation
and are mostly contained within the 
$\pm$1$\sigma$ band constructed from simulated data sets with
statistical and systematic fluctuations. 
The high-precision data at multiple baselines allow exclusion of a
large section of ($\sin^22\theta_{14}$, $|\Delta m^{2}_{41}|$)
parameter space. 
The sensitivity in the $0.01 \lesssim |\Delta m^{2}_{41}| \lesssim 0.3~{\rm eV}^2$ region originates predominantly from the relative spectral comparison between the two near halls, and in the $|\Delta m^2_{41}| \lesssim 0.01~{\rm eV}^2$ region from the comparison between the near and far halls.
The dip structure at $|\Delta m^{2}_{41}| \approx |\Delta m^{2}_{32}| \approx 2.4 \times
10^{-3}~{\rm eV}^2$ is due to
the degeneracy between $\sin^{2}2\theta_{14}$ and $\sin^{2}2\theta_{13}$.
The fine structure of the data contours compared to the expectation 
originates from statistical fluctuations in the data.

In Figure~\ref{fig:excludeOsc}, there is a slight difference between 
the CL contour from method A and the ${\rm CL_{s}}$ contour
from method B for $|\Delta m^2_{41}| \lesssim 2 \times 10^{-3}~\rm{eV}^{2}$. 
In this region, most of the oscillation effects appear in the far hall at prompt energies $\lesssim2$~MeV, where the statistics are more limited.
A study based on a large sample of Monte Carlo pseudo-experiments determined that the two methods react differently to statistical fluctuations and produce slightly different limits in this region. The difference observed in Figure~\ref{fig:excludeOsc} is found to be consistent with the expectation from this study at the $\sim 1\sigma$ level. 

The resulting limits on $\sin^22\theta_{14}$ are improved by roughly
a factor of 2 compared to the previous
publication~\cite{An:2014Sterile}. The increased statistics are the largest contributor to this
improvement, although the reductions in background and in the AD-uncorrelated
energy scale uncertainty also play a role.  
The uncertainty in $|\Delta m^2_{32}|$ is the dominant systematic
uncertainty in the $|\Delta m^2_{41}| \lesssim |\Delta m^2_{32}|$ region,
while for higher values of $|\Delta m^2_{41}|$ the AD-uncorrelated energy scale and detector efficiency uncertainties are dominant.
The total uncertainty is dominated by the statistics; another factor of 2 improvement in sensitivity is expected by 2017.
This result can be combined with $\pbar{\nu}_\mu$ disappearance searches~\cite{MinosPaper} 
 in order to constrain $\pbar{\nu}_\mu\to \pbar{\nu}_e$ transitions~\cite{CombinedPaper}, since the oscillation probability of $\pbar{\nu}_\mu \to \pbar{\nu}_e$ 
in the four-neutrino scenario is 
approximately proportional to $|U_{e4}|^2|U_{\mu 4}|^2$, 
and the individual sizes of $|U_{e4}|^2$ and $|U_{\mu4}|^2$ can be
constrained with $\pbar{\nu}_e$ and $\pbar{\nu}_\mu$ disappearance searches, respectively.

In summary, we report an improved search for light sterile neutrino mixing 
with the full configuration of the Daya Bay Reactor Neutrino Experiment in the electron antineutrino disappearance channel. 
No evidence of a light sterile neutrino is found through a relative 
comparison of the observed antineutrino energy spectra at the
  three experimental halls.
With 3.6 times the statistics of the previous publication, 
these results set the most stringent limits to date on $\sin^22\theta_{14}$ in the $2\times10$$^{-4} \lesssim |\Delta m^{2}_{41}| \lesssim 0.2$ eV$^{2}$ region. 

The Daya Bay Reactor Neutrino Experiment is supported in part by the Ministry of Science and Technology of China, the U.S. Department of Energy, the Chinese Academy of Sciences (CAS), the CAS Center for Excellence in Particle Physics, the National Natural Science Foundation of China, the Guangdong provincial government, the Shenzhen municipal government, the China General Nuclear Power Group, the Laboratory Directed Research and Development Program of the Institute of High Energy Physics, the Shanghai Laboratory for Particle Physics and Cosmology, the Research Grants Council of the Hong Kong Special Administrative Region of China, the University Development Fund of The University of Hong Kong, the MOE program for Research of Excellence at National Taiwan University, National Chiao-Tung University, MOST funding support from Taiwan, the U.S. National Science Foundation, the Alfred P. Sloan Foundation, the Laboratory Directed Research and Development Program of Berkeley National Laboratory and Brookhaven National Laboratory, the Ministry of Education, Youth, and Sports of the Czech Republic, Charles University in Prague, the Joint Institute of Nuclear Research in Dubna, Russia, the NSFC-RFBR joint research program, and the National Commission for Scientific and Technological Research of Chile. We acknowledge the Yellow River Engineering Consulting Co., Ltd., and the China Railway 15th Bureau Group Co., Ltd., for building the underground laboratory. We are grateful for the ongoing cooperation from the China General Nuclear Power Group and China Light and Power Company.

\bibliographystyle{apsrev4-1}

\bibliography{sterile}

\end{document}